\documentclass[preprint,prd,amsmath,nofootinbib,twoside]{revtex4}
\usepackage{braket}
\usepackage{graphicx}
\usepackage{hyperref}

\begin{document}
\title{Disentangling Neutrino Oscillations}
\date{\today}

\author{Andrew G. Cohen}
\email{cohen@bu.edu}
\author{Sheldon L. Glashow}
\email{slg@bu.edu}
\affiliation{Physics Department,  Boston University\\
  Boston, MA 02215, USA}
\author{Zoltan Ligeti}
\email{ligeti@lbl.gov}
\affiliation{Ernest Orlando Lawrence Berkeley National Laboratory\\
  University of California, Berkeley, CA 94720}

\begin{abstract}
  The theory underlying neutrino oscillations has been described at
  length in the literature. The neutrino state produced by a weak
  decay is usually portrayed as a linear superposition of mass
  eigenstates with, variously, equal energies or equal momenta. We
  point out that such a description is incorrect, that in fact, the
  neutrino is entangled with the other particle or particles emerging
  from the decay. We offer an analysis of oscillation phenomena
  involving neutrinos (applying equally well to neutral mesons) that
  takes entanglement into account. Thereby we present a theoretically
  sound proof of the universal validity of the oscillation formul\ae\
  ordinarily used.  In so doing, we show that the departures from
  exponential decay reported by the GSI experiment cannot be
  attributed to neutrino mixing. Furthermore, we demonstrate that the
  `M\"ossbauer' neutrino oscillation experiment proposed by Raghavan,
  while technically challenging, is correctly and unambiguously
  describable by means of the usual oscillation formal\ae{}.
\end{abstract}

\maketitle{}

\section{Introduction}

Neutrino oscillations are among the most interesting phenomena
discovered in particle physics in recent years. Although these
oscillations were anticipated long ago
\cite{Pontecorvo:1967fh,Gribov:1968kq}, their detection was
complicated by the small size of the neutrino masses. Today, however,
oscillation phenomena have been observed and studied for neutrinos
originating from the sun, nuclear reactors, accelerators, and
cosmic-ray interactions in the atmosphere.  For a review see
\cite{Bilenky:2004xm}.

Recently, several novel and ingenious experiments have been suggested
(and in at least one case carried out) to further explore the physics
of neutrino masses.  Raghavan has proposed the study of oscillations
via the resonant capture of anti-neutrinos produced by the bound-state
beta decay of tritium
\cite{Raghavan:2005gn,Raghavan:2006xf,Raghavan:2008cs,Raghavan:2008tb}. This
suggestion has led to some confusion.  Akhmedov et
al.~\cite{Akhmedov:2008jn,Akhmedov:2008zz} agree that oscillations
should be expected in this experiment, whilst Bilenky et
al.~\cite{Bilenky:2008ez,Bilenky:2008dk} conclude that whether or not
oscillations are seen can ``test fundamentally different approaches to
neutrino oscillations''.

In addition, Litvinov et al.~\cite{Litvinov:2008rk} report the
observation of non-exponential weak decays of hydrogenic ions.  Some
theoretical analyses interpret these data in terms of neutrino
mixing~\cite{Lipkin:2008ai,Ivanov:2008sd,Kleinert:2008ps,Lipkin:2008in}
while others refute such an
interpretation~\cite{Giunti:2008ex,Peshkin:2008vk,Kienert:2008nz,Gal:2008sw}.
In another experiment a stronger bound was set~\cite{Vetter:2008ne} on
the amplitude of the oscillatory modulation of the exponential decay
of $^{142}$Pm at the frequency reported in~\cite{Litvinov:2008rk}.

Our motivation for this work is to produce a simple and coherent
theoretical framework for describing oscillation experiments involving
elementary particles. Although a proper treatment of oscillation
phenomena may appear (implicitly) in the literature\footnote{For
  example, the work of Nauenberg~\cite{Nauenberg:1998vy}, while not
  identical to our approach, correctly identifies entanglement as
  necessary for energy--momentum conservation. Similarly
  Kayser~\cite{Kayser:1981ye} recognized that sufficiently accurate
  momentum measurements prevent oscillations.}, the significant
discrepancies and imprecisions in existing approaches to neutrino
oscillations suggest the need for such a unified framework.

We use neutrinos as our primary example in the derivation of the
oscillation formul\ae{}. As we shall see, our results apply equally
well to other types of elementary particle oscillations including
those of $B$ and $K$ mesons. We discuss neutral meson oscillations in
\ref{sec:neut-meson}.

\section{Universal Oscillations}
\label{sec:univ-oscill}

Neutrino oscillations arise because the weak interactions conserve
lepton flavor whereas energy eigenstate neutrinos are not flavor
eigenstates.  Most analyses describe the production of neutrinos (via
a weak decay or scattering event) in terms of a flavor eigenstate
which is then decomposed as a linear combination of mass eigenstate
neutrinos, each of which propagates according to its own dispersion
relation. Often an analogy is drawn with a simple two state system
(for ease of notation we restrict to two neutrino flavors with mixing
angle $\theta$; the generalization to three flavors is
straightforward), and frequently one sees formul\ae{} like
\begin{equation}
  \label{eq:1}
  \ket{\nu_{e}} = \cos\theta \ket{\nu_{L}}+\sin\theta \ket{\nu_{H}}
\end{equation}
where we have labeled the mass eigenstates as ``H''eavy and
``L''ight. This approach is not entirely correct and has led to
significant confusion in the literature.  For example, the states must
depend on the three-momenta of the neutrinos. But because $\nu_{L}$ and
$\nu_{H}$ have different masses it is not possible for this
superposition to be an eigenstate of both energy and momentum, thus
leading some authors to suggest a common energy while others prefer a
common momentum. However neither of these suggestions can be correct,
because neither can account for simultaneous energy and momentum
conservation in the weak process that produces the neutrino.

The resolution to this puzzle is quite simple: the state produced
following the weak interaction is \emph{not\/} of the form
\eqref{eq:1}. Rather, the state produced has the neutrino mass
eigenstates entangled with the other particles remaining after the
weak process has occurred. Energy and momentum are fully conserved by
the process, as must be the case given space-time translation invariance
of the underlying interaction.

A simple example serves to illustrate the primary issues. Consider a
particle $N$ (the ``parent'') of mass $M$ which decays to another
particle $n$ (the ``daughter'') of mass $M^{\prime}$ plus a
neutrino.\footnote{The example of a 2-body decay exhibits all the
  features of interest, and extension to other processes requires no
  significant modifications. The particle could be a pion decaying
  conventionally to a muon or equally well an atom decaying via
  electron capture.} To simplify our discussion we ignore the spins
of all particles involved as well as any internal excitations.  By
assuming the parent to be sufficiently long-lived, we may choose the
initial state to have arbitrarily well-defined energy and momentum $P$
and we may treat the decay process in perturbation theory. In this
approximation we may think of the decay as occurring instantaneously
at some time (distributed in accord with the exponential decay law)
leaving us in the state
\begin{equation}
  \label{eq:2}
  \ket{\psi} =\frac{1}{\sqrt{\mathcal{N}}} \Bigl[
  \int\!D_{2}(k_{l},q_{l})\,\,\cos\theta \Ket{n(k_{l}) \nu_{L}(q_{l})} +
  \int\!D_{2}(k_{h},q_{h})\,\,\sin\theta \Ket{n(k_{h}) \nu_{H}(q_{h})}
  \Bigr]
\end{equation}
where $q_{i}^{2} = m_{i}^{2}$ and $k_{i}^{2} = {M'}^{2}$.  The
phase space for the two particles $D_{2}(k,q)$ is
\begin{equation}
  \label{eq:3}
  D_{2}(k,q) = \frac{d^{3}k}{(2\pi)^{3}2 E_{k}}
  \frac{d^{3}q}{(2\pi)^{3}2 E_{q}} (2\pi)^{4} \delta^{4}(P-k-q)
\end{equation}
where the energies $E_{k}, E_{q}$ are computed with the appropriate
particle masses and, for simplicity, we have assumed an amplitude
independent of momenta.  The value of the normalization constant
$\mathcal{N}$ will not be needed.\footnote{For reference $\mathcal{N}
  = VT \cdot 2 M \Gamma$ where $VT$ is the volume of space-time and
  $\Gamma$ is the parent particle decay rate.}  Note that all particles
are on the mass-shell and $\ket\psi$ is an eigenstate of energy and
momentum with eigenvalue $P$.  This is achieved through the
entanglement of the neutrino with the daughter particle and would not be
possible if the state were a non-entangled product with the ket of
\eqref{eq:1} as a factor.

The latter point is worth emphasizing. Flavor-charge operators, such
as the electron or muon number operators, remain well-defined in the
Standard Model augmented with neutrino mixing but no longer commute
with the Hamiltonian. The lepton flavor conserving weak interactions
are most simply written in terms of the electron (muon) neutrino field
with definite flavor which acts on a state so as to alter the electron
(muon) number by one unit. However, since time evolution alters the
flavor, it is not very fruitful to consider states of definite
flavor. Rather, although the \emph{fields\/} that create and
annihilate mass eigenstates are formed as linear combinations of the
\emph{fields\/} of definite flavor, the corresponding construction for
\emph{states\/} is not helpful. This situation is much like the
relation between \emph{chirality\/} (a useful property of fields) and
\emph{helicity\/} (a measurable property of states).

Having properly identified the final state, how are we to treat
oscillations?  Most oscillation experiments observe the neutrino as it
produces a charged lepton via a weak interaction, and ignore any other
particles that accompany the neutrino's production. Because the
neutrino is entangled with these other (undetected) particles, we must
construct the density matrix for the neutrino by tracing over these
other degrees of freedom. Neutrino oscillations arise from an
off-diagonal term in this density matrix of the form
$\ket{\nu_{L}}\bra{\nu_{H}}$. Constructing the density matrix from the
state \eqref{eq:2}, we obtain
\begin{multline}
  \label{eq:4}
  \rho_{\nu} = \frac{1}{\sqrt{\mathcal{N}}} \Bigl[
  \int D_{2}(k_{l},q_{l}) D_{2}(\tilde{k}_{l},
  \tilde{q}_{l}) \cos^{2}\theta \Braket{n(k_{l})|n(\tilde{k}_{l})}
  \Ket{\nu_{L}(q_{l})}\Bra{\nu_{L}(\tilde{q}_{l})} \\
  + \int D_{2}(k_{l},q_{l}) D_{2}(\tilde{k}_{h},
  \tilde{q}_{h}) \cos\theta\sin\theta \Braket{n(k_{l})|n(\tilde{k}_{h})}
  \Ket{\nu_{L}(q_{l})}\Bra{\nu_{H}(\tilde{q}_{h})} + \text{h.c.}\\
  + \int
  D_{2}(k_{h},q_{h}) D_{2}(\tilde{k}_{h}, \tilde{q}_{h})
  \sin^{2}\theta \Braket{n(k_{h})|n(\tilde{k}_{h})}
  \Ket{\nu_{H}(q_{h})}\Bra{\nu_{H}(\tilde{q}_{h})} 
  \Bigr]\ .
\end{multline}
However the cross terms between $\ket{\nu_{L}}$ and $\ket{\nu_{H}}$ on
the middle line vanish.  A non-zero inner product for the daughter
particle ($E^{\prime}_{k} \equiv \sqrt{\mathbf{k}^{2}+{M'}^{2}}$)
\begin{equation}
  \label{eq:5}
  \Braket{n(k_{l})|n(\tilde{k}_{h})} = (2\pi)^{3} 2E^{\prime}_{k_{l}}
  \delta^{3}(\mathbf{k}_{l} - \tilde{\mathbf{k}}_{h})
\end{equation}
requires that the two momenta be equal, while the delta functions in
$D_{2}$ reflecting energy--momentum conservation require that $k_{l} -
\tilde{k}_{h} = \tilde{q}_{h}-q_{l}$. But the two neutrino states have
different invariant masses and so this momentum difference can never
vanish.  Hence these daughter particle states are orthogonal and the neutrino
density matrix is diagonal
\begin{equation}
  \label{eq:6}
  \rho_{\nu} \propto  \int \frac{D_{2}(k_{l}, q_{l})}{2 E^{(\nu_{L})}_{q_{l}}}
  \Ket{\nu_{L}(q_{l})}\Bra{\nu_{L}(q_{l})}\cos^{2}\theta  + 
  \int \frac{D_{2}(k_{h}, q_{h})}{2 E^{(\nu_{H})}_{q_{h}}}
  \Ket{\nu_{H}(q_{h})}\Bra{\nu_{H}(q_{h})} \sin^{2}\theta
\end{equation}
with probability $\cos^{2}\theta$ of containing $\nu_{L}$ and
probability $\sin^{2}\theta$ of containing $\nu_{H}$.  Since the
amplitude for the detection of $\nu_{L}$ via an electron-implicated
weak interaction is $\cos\theta$ and that for $\nu_{H}$ is
$\sin\theta$, this leads to a detection probability proportional to
$\cos^{4}\theta+\sin^{4}\theta$, exactly as we expect in the absence
of interference.  When the decay products of an initial state of
well-defined momentum evolve without further interaction no
oscillation phenomena appear.

So how can neutrino oscillations arise? The assumptions of the final
sentence of the preceding paragraph must not apply to experiments
that exhibit oscillations. In fact, so long as the neutrino remains
entangled as in \eqref{eq:2}, there is no possibility of interference
and hence no possibility of oscillation.  To realize oscillations the
neutrino mass eigenstates must be disentangled.

We have so far treated the parent particle as an exact energy and
momentum eigenstate with an associated unrealistic uniform detection
probability throughout spacetime.  This is surely not the case in
realistic circumstances. Nonetheless, it is instructive to consider
this unrealistic state in the situation where the daughter particle is
detected in addition to the neutrino. For neutrinos produced in pion
decay, for example, the associated muon (or its decay products) might
be detected in a state $\Ket{\bar{n}}$.  Rather then tracing over the
unobserved daughter this case requires computation of the
joint probability for observation of the daughter in the state
$\Ket{\bar{n}}$ along with the neutrino.  This may be calculated by
projecting the state \eqref{eq:2} by
$\Ket{\bar{n}}\Bra{\bar{n}}$. This projection then disentangles the
state \eqref{eq:2}, leaving the neutrino in a simple superposition.
The neutrino itself is unaffected by this projection: the two
components continue to have the momenta $q_{l}, q_{h}$ determined by
the decay kinematics.

This projection alters the amplitude of the $\nu_{L}$ and $\nu_{H}$
components in the superposition by the two matrix elements
$\Braket{\bar{n}\vert n(k_{l,h})}$. The state $\Ket{\bar{n}}$ is
typically well-localized in space-time, and hence has a substantial
spread in momentum.  Because the momenta $k_{l,h}$ are nearly the same
the matrix elements $\Braket{\bar{n}\vert n(k_{l})}\text{ and
}\Braket{\bar{n}\vert n(k_{h})}$ are, for all practical purposes,
equal. Hence, subsequent to this projection the neutrino may be treated as
a superposition of the two mass eigenstates (as is usually done) with
momenta $q_{l}\text{ and }q_{h}$:
\begin{equation}
  \label{eq:7}
  \ket\psi \sim \cos\theta \Ket{\nu_{L}(q_{l})} +
  \sin\theta\Ket{\nu_{H}(q_{h})} \ .
\end{equation}
We have restricted the superposition to one spatial dimension,
eliminating the integral over the neutrino direction. This is a
reasonable approximation because oscillation experiments require the
neutrino to propagate far from the production point, hence we detect
only those particles traveling in the appropriate direction.  As
promised in the introduction, the neutrinos are neither equal in
energy nor equal in momentum. The detection of the neutrino may be
modeled by acting with an operator of electron flavor at the detector
space-time location $z\equiv(t, d)$ (as usual we work in the Heisenberg
picture) giving a detection amplitude
\begin{equation}
  \label{eq:8}
  \mathcal{A} \sim \cos^{2}\!\theta\, e^{i q_{l}\cdot z} +
  \sin^{2}\!\theta \,e^{i  q_{h}\cdot z}\ .
\end{equation}
The square of this expression contains an interference term between
the $H$ and $L$ amplitudes which may produce oscillations. Although
the amplitudes in \eqref{eq:8} show only complex-exponential
dependence on the detection location, realistic experiments involve
amplitudes that have an extended space-time support localized around
the trajectory $d = v t$. The $H$ and $L$ amplitudes interfere only
when they have common support.  Because the particles have velocity
dispersions with slightly different central values, they separate
as they travel towards the detection event. Interference is possible
only if this separation is smaller than the localization size of the
particle $v \Delta T$, or what is often called the size of the
wave-packet.  The condition for interference is
\begin{equation}
  \label{eq:9}
  \Bigl|\frac{v_{h}-v_{l}}{v_{h}+v_{l}}\Bigr| =
  \Bigl|\frac{\sigma\omega\delta  q-\delta\omega\sigma
    q}{\sigma\omega\sigma q - \delta\omega\delta  q}\Bigr|  \ll
  \frac{\Delta T}{t}
\end{equation}
where the sum and difference of the neutrino energies are
$\sigma\omega\equiv \omega_{h}+\omega_{l},\delta \omega \equiv
\omega_{h}-\omega_{l}$ and $\sigma q, \delta q$ are the corresponding
expressions for the sum and difference of the magnitudes of the
spatial momenta.

The interference term in the square of the amplitude \eqref{eq:8} has
the phase $\phi \equiv (q_{h}-q_{l})\cdot z$. So far we have made no
assumptions about the masses of the particles involved, nor about the
momentum of the initial parent that gives rise to the neutrino. This
generality allows us to describe oscillations of other particles (such
as $K$ and $B$ mesons) as well as neutrinos. The only assumption we
make at this stage is that the difference in velocities between the
two components is small enough so that the particles may interfere in
the detector located at $z\equiv(t,d)$: Eq. \eqref{eq:9}. This
condition applies to $K$ meson oscillations, $B$ meson oscillations
and neutrino oscillations under all realistic conditions. We continue
to refer to the oscillating particles as neutrinos in the sequel.

Condition \eqref{eq:9} ensures that the two components of the state
overlap at the detection point, thus allowing them to interfere.  For
reasonable velocity dispersions this overlap may be evaluated using
stationary phase and is dominated when neutrino velocities are $v =
d/t$.  Thus we may take $\sigma q/\sigma\omega=d/t$.  Provided we
observe the neutrinos over times such that the two components have not
spatially separated, the space-time vector $z \equiv (t,d)$ may then
be expressed as
\begin{equation}
  \label{eq:10}
  z = (t, d) \simeq t\left(1,\frac{\sigma q}{\sigma\omega}\right) =
  \frac{t}{\sigma\omega}
  \left(q_{h}+q_{l}\right)\ .
\end{equation}
The oscillation phase is then
\begin{equation}
  \label{eq:11}
  \phi \equiv (q_{h}-q_{l})\cdot z =
  \frac{t}{\sigma\omega}(q_{h}-q_{l})\cdot (q_{h}+q_{l}) = t 
  \frac{\delta m^{2}}{\sigma\omega} \ .
\end{equation}
This is the usual answer for relativistic neutrinos where $t\simeq d$
and $\sigma\omega$ is just twice the neutrino energy. But the same
expression applies whenever \eqref{eq:9} is satisfied, relativistic or
not. For non-relativistic particles, for example, we have
$\sigma\omega \simeq m_{l}+m_{h}$ and the phase $\phi$ is then $t
\delta m$.

In this argument we used no properties of the vectors $q_{l,h}$ other
than the condition \eqref{eq:9}.  The energies and magnitudes of the
spatial momenta are fully determined: the two neutrinos are neither
equal in energy nor momentum. Nevertheless, use of such incorrect
values fortuitously leads to the correct oscillation phase.

Moreover, the detailed properties of the state $\ket{\bar{n}}$, other
than the near equality of the matrix elements $\Braket{\bar{n}\vert
  n(k_{l,h})}$, played no role in our analysis. Similarly the mechanism
producing the neutrino and any distribution in its momentum are
irrelevant to \eqref{eq:11} provided \eqref{eq:9} is satisfied. In
this sense the oscillation phase of \eqref{eq:11} is ``universal''.

Usually the neutrinos are detected without accompanied detection of
the daughter particle. In this case we must employ the density matrix
for the neutrino after tracing over the daughter Hilbert space.
Interference between the $\nu_{L}$ and $\nu_{H}$ components requires a
non-vanishing inner product of the daughter states in this trace. In
practice this is realized by accounting for a momentum spread arising
from the parent.  For any realistic experiment, the parent state is
not a momentum eigenstate but rather a superposition of momenta in a
narrow range. If this range is such that the daughter particle
accompanying $\nu_{L}$ can have the same four-momentum as the daughter
particle accompanying $\nu_{H}$, then oscillations become possible.
%The finite lifetime of the unstable parent particle also produces a
%spread in momentum with a similar effect. 
The difference between the daughter particle momenta in the two
components is of order $\delta m^{2}/\sigma\omega$.  For realistic
neutrino masses and energies, the required momentum difference is
exceedingly small, less than $10^{-10}$ eV.  Because realistic
experiments always start with an initial state at least slightly
localized in space-time (often to a nuclear distance, but surely to
within a kilometer or better) this momentum difference always lies
within the initial momentum spread.

As an example consider the long-lived parent particle as above but in
an initial state which is a superposition of spatial momenta in a
narrow band.\footnote{The finite lifetime of the unstable initial
  particle produces an additional (Lorentzian) spread in the invariant
  mass of the daughter plus the neutrino. This effect may be
  incorporated similarly to our inclusion of the spatial momentum
  spread.} This may be described by superposing slight boosts
$\Lambda_{v}$ of the initial particle momentum $P$ in the state
\eqref{eq:2}.  Assuming the initial momentum spread is small
corresponds to requiring that only $v \ll 1$ appears in this
superposition.  The final state \eqref{eq:2} is then replaced by a
similar superposition
\begin{multline}
  \label{eq:12}
  \ket{\psi} =\frac{1}{\sqrt{\mathcal{N}}} \int d^{3}\!v
  f(\mathbf{v}) \Bigl[
  \int\!D_{2}(k_{l},q_{l})\,\,\cos\theta \Ket{n(\Lambda_{v}k_{l})
    \nu_{L}(\Lambda_{v}q_{l})} \\ + 
  \int\!D_{2}(k_{h},q_{h})\,\,\sin\theta \Ket{n(\Lambda_{v}k_{h})
    \nu_{H}(\Lambda_{v}q_{h})} 
  \Bigr]
\end{multline}
where $f(\mathbf{v})$ describes the superposition and we have used the
Lorentz invariance of the phase space.  We may simplify as before by
restricting to a single direction
\begin{equation}
  \label{eq:13}
  \ket\psi \sim \int dv f(v)
  \Bigl[
  \cos\theta \Ket{n(\Lambda_{v}k_{l}) \nu_{L}(\Lambda_{v}q_{l})} +
  \sin\theta \Ket{n(\Lambda_{v}k_{h}) \nu_{H}(\Lambda_{v}q_{h})}
  \Bigl]
\end{equation}
where $k_{i} \text{ and } q_{i}$ are now completely determined in
terms of the initial momentum $P$ by conservation of energy and
momentum. Although not obvious at first glance, this superposition
allows for the disentanglement of the neutrino mass from the daughter
momentum.

To see this, note that the vectors $k_{l}$ and $k_{h}$ have the same
invariant mass, and hence differ from a common four-momentum $k$ by
(small) Lorentz boosts: $\Lambda_{v_{0}}k_{h} = \Lambda_{-v_{0}}
k_{l}\equiv k$. The boost velocity $v_{0}$ is easily computed in terms
of $q_{l} \text{ and } q_{h}$:
\begin{equation}
  \label{eq:14}
  v_{0} = - \frac{\delta q}{2E_{p}-\sigma\omega} \ .
\end{equation}
By shifting the velocity $v$ in the integrals in the two terms of
\eqref{eq:13} relative to each other we can rewrite the superposition
as
\begin{equation}
  \label{eq:15}
  \ket\psi \sim \int dv \Ket{n(\Lambda_{v}k)}\otimes
  \Bigl[
  f(v+v_{0})\cos\theta \Ket{\nu_{L}(\Lambda_{v}\Lambda_{v_{0}}q_{l})} + 
  f(v-v_{0})\sin\theta \Ket{\nu_{H}(\Lambda_{v}\Lambda_{-v_{0}}q_{h})} 
  \Bigl].
\end{equation}
Although this state is still entangled (a sum of products), the
neutrino \emph{mass\/} is not fully entangled with the daughter
momentum.  The density matrix for the neutrino constructed upon
tracing over the daughter states now contains cross terms between
neutrino $L$ and neutrino $H$ which give rise to oscillations.

Our previous calculation of the oscillation phase continues to apply,
subject to two changes. Firstly the interference term contains a
factor of $f(v+v_{0}) f^{*}(v-v_{0})$ rather than the $\vert
f(v+v_{0})\vert^{2} \text{ or } \vert f(v-v_{0})\vert^{2}$ factors of
the diagonal terms. If, as is generally the case, the function $f$
representing the momentum superposition of the parent does not vary
significantly on the scale of the small velocity $v_{0}$, these
factors are all essentially the same.  Secondly the energy
$\sigma\omega$ that appears is modified by the boosts:
\begin{equation}
  \label{eq:16}
  \sigma\omega \to \sigma\omega - v_{0} \delta q + v \sigma q \ .
\end{equation}
If the support of $f$ is such that $v \ll 1$, so that the momentum of
the parent is moderately well defined, we may drop the terms
proportional to $v$ and $v_{0}$. Once again we obtain the familiar
oscillation formula, and once again the details of the momentum
superpositions involved play no role other than ensuring the presence
of the interference term in the neutrino density matrix.

We conclude this section with a brief discussion of the novel
oscillation-related experiments mentioned earlier.  Consider first the
proposal of
Raghavan~\cite{Raghavan:2006xf,Raghavan:2005gn,Raghavan:2008cs,Raghavan:2008tb}
to study the resonant capture of antineutrinos from bound-state
tritium decay. The question of whether or not such ``M\"ossbauer
neutrino oscillations'' are present has been hotly contested. Bilenky
et al.~\cite{Bilenky:2008ez,Bilenky:2008dk} conclude that such
oscillations may or may not occur and that the Raghavan experiment
``provides the unique possibility to discriminate  basically
different approaches'' to neutrino oscillations.  Contrariwise,
Akhmedov et al.~\cite{Akhmedov:2008zz} find that ``a proper interpretation of the time-energy
uncertainty relation is fully consistent with oscillations of
M\"ossbauer neutrinos.'' The result of our analysis is
simple. Condition \eqref{eq:9}, our unique and simple criterion for
the appearance of oscillations is satisfied by the Raghavan
experiment: if the Raghavan experiment can be realized, it will be a
powerful tool with which to study neutrino oscillations.  Furthermore
and contrary to Bilenky et al., there is no ambiguity about the
approach to neutrino oscillations for the Raghavan experiment to
resolve.

\begin{figure}
  \centering
  \includegraphics[width=10cm]{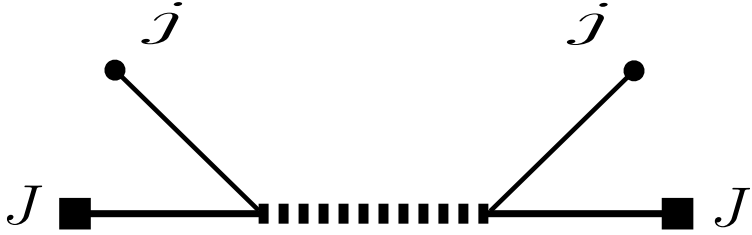}
  \caption{Feynman diagram representing  $|\mathcal{A}|^{2}$ for
    the observation of the daughter particle. The square blobs represent
    the parent source, the round blobs the daughter detector, and the
    dashed line is the (on-shell) neutrino.}
  \label{fig:1}
\end{figure}

Now let us turn to the GSI experiment. An essential feature of this
experiment is that the neutrino is not detected: the observed
oscillations appear in measurements of the time of disappearance of
the parent particle (coincident with the appearance of the daughter
particle). As shown below, the arguments we have adduced demonstrate
that experiments which do not observe the neutrino cannot display
interference. Our discussion so far has not included the production
and decay of the parent particle, but this is easily incorporated. We
create the parent particle by acting on the vacuum with some (smeared)
operator $N^{\dagger}_{J}$ producing the parent around $t=0$, and
model the observation of the daughter at a subsequent time by acting
with a (smeared) operator $n_{j}$ around the time $t$.  The neutrino
is not observed and remains in the final state.  The squared amplitude
for this process is represented diagrammatically in
Fig.~\ref{fig:1}. An on-shell neutrino in the final state is
represented by the dashed line, corresponding to a cut propagator
$\delta (q^{2}-m^{2}_{i})\theta(q^{0})$. The full squared amplitude is
given by a sum over the several neutrino mass eigenstates. Notice that
there are no cross terms between these mass eigenstates: because the
neutrinos have different invariant masses, there is no possibility of
interference between them.
Thus the reported oscillation of the decay time cannot be explained in
terms of  interference between neutrino states.

\section{Neutral Meson Mixing}
\label{sec:neut-meson}

We turn to the study of mixing effects of mesons, in particular in the
$B$ meson sector.  Our discussion is framed to most closely resemble
the experiments conducted at $B$ factories, but a similar analysis
applies to other cases.  Measurements at the $B$ factories observe the
$B$ mesons produced in the decay of the $\Upsilon(4S)$.  We consider
the case where one of the mesons is detected at a space-time location
$z$ through its decay to the state $\ket{\mathcal{O}}$ and the other
at space-time location $\tilde{z}$ through its decay to the state
$\ket{\tilde{\mathcal{O}}}$. The entangled state resulting from the
decay of an $\Upsilon(4S)$ into neutral $B$ mesons is
\begin{equation}
  \label{eq:17}
  \Ket\psi = \frac{1}{\sqrt2} \Bigl[\Ket{B_L(k_{l}) B_H(\tilde{k}_{h})} -
  \Ket{B_H(k_{h}) B_L(\tilde{k}_{l})}\Bigr]
\end{equation}
where $L,H$ label the light and heavy mass eigenstates and, as in the
previous section, we have restricted the momenta to those pointing in
the directions of the observation events. By convention the momenta
without over-tildes point toward the event $z$ while those with
tildes point toward $\tilde{z}$.

The observations of the $B$ mesons may be modeled by taking the matrix
element of an appropriate local operator that annihilates the
particles in the final states $\ket{\mathcal{O}},\ket{\tilde{\mathcal{O}}}$ between $\ket\psi$ and the vacuum:
\begin{multline}
  \label{eq:18}
  S_{\mathcal{O}\tilde{\mathcal{O}}} = \Braket{0|\mathcal{O}(z)
    \tilde{\mathcal{O}}(\tilde{z})|\psi} = \frac{1}{\sqrt{2}} \Bigl[
  \braket{0|\mathcal{O}(z)|B_{L}(k_{l})}
  \braket{0|\tilde{\mathcal{O}}(\tilde{z})| B_{H}(\tilde{k}_{h})} \\ -
  \braket{0|\mathcal{O}(z)|B_{H}(k_{h})}
  \braket{0|\tilde{\mathcal{O}}(\tilde{z})| B_{L}(\tilde{k}_{l})}
  \Bigr]\ .
\end{multline}
The space-time dependence of these matrix elements is determined by
translation invariance:
\begin{equation}
  \label{eq:19}
  \begin{split}
  \braket{0|\mathcal{O}(z)|B_{L}(k_{l})} =& e^{ik_{l}\cdot z}
  \braket{0|\mathcal{O}(0)|B_{L}(k_{l})} \equiv e^{ik_{l}\cdot z}
  A_{L} \\
  \braket{0|\tilde{\mathcal{O}}(\tilde{z})|B_{L}(\tilde{k}_{l})} =&
  e^{i\tilde{k}_{l}\cdot \tilde{z}} 
  \braket{0|\tilde{\mathcal{O}}(0)|B_{L}(\tilde{k}_{l})} 
  \equiv e^{i\tilde{k}_{l}\cdot \tilde{z}}
  \tilde{A}_{L}
  \end{split}
\end{equation}
and similarly for the matrix elements involving $B_{H}$.  

Taking the absolute-value squared of $S$, we obtain
\begin{multline}
  \label{eq:20}
  \vert S_{\mathcal{O}\tilde{\mathcal{O}}}\vert^{2} = \frac{1}{2}
  \Bigl\{
    \vert A_{L} \vert^{2}\vert \tilde{A}_{H} \vert^{2}
    e^{i(k_{l}-k^{*}_{l})\cdot z } e^{i(\tilde{k}_{h}-\tilde{k}^{*}_{h})\cdot \tilde{z}} \\
    - A_{L}A_{H}^{*}\tilde{A}_{H}\tilde{A}_{L}^{*}
    e^{i(k_{l}-k_{h}^{*})\cdot z }
    e^{i(\tilde{k}_{h}-\tilde{k}_{l}^{*})\cdot \tilde{z}} -
    \text{c.c.} \\
    + \vert A_{H} \vert^{2}\vert \tilde{B}_{L} \vert^{2} 
        e^{i(k_{h}-k^{*}_{h})\cdot z }
        e^{i(\tilde{k}_{l}-\tilde{k}^{*}_{l})\cdot \tilde{z}}  
   \Bigr\}\ .
\end{multline}
We have uncharacteristically kept the complex conjugation on the
momenta of the $B$ mesons. This is to keep track of the finite
lifetime of the mesons that may be incorporated as an imaginary part
for the energy.\footnote{We are being slightly sloppy. A proper
  treatment would use a local operator to create the $B$ meson from
  the vacuum and then follow its propagation. For a width small
  compared to the mass this propagator is dominated by a simple pole
  that is not on the real axis but rather on the second sheet, with an
  imaginary part given by the decay width. The net result is the
  complex exponential in \eqref{eq:20}.} Using the formula derived in
the previous section, we have
\begin{equation}
  \label{eq:21}
  \begin{split}
    i(k_{l}-k_{l}^{*})\cdot z  = -\Gamma_{L} t \qquad
    i(k_{h}-k_{h}^{*})\cdot z  = -\Gamma_{H} t  \\
    i(\tilde{k}_{l}-\tilde{k}_{l}^{*})\cdot \tilde{z} =
    -\tilde{\Gamma}_{L} \tilde{t}\qquad 
    i(\tilde{k}_{h}-\tilde{k}_{h}^{*})\cdot \tilde{z} =
    -\tilde{\Gamma}_{H} \tilde{t} \\ 
    i(k_{l}-k_{h}^{*})\cdot z  = -\frac{\Gamma_{L}+\Gamma_{H}}{2} t - i t
    \frac{\delta m^{2}}{\sigma\omega} \\
    i(\tilde{k}_{h}-\tilde{k}_{l}^{*})\cdot \tilde{z} =
    -\frac{\tilde{\Gamma}_{L}+\tilde{\Gamma}_{H}}{2} \tilde{t} + i
    \tilde{t} \frac{\delta  m^{2}}{\sigma\tilde{\omega}}  \ .
  \end{split}
\end{equation}
In the laboratory frame the two mesons generally have (slightly)
different velocities. For completeness, we have kept the
difference between the widths $\Gamma, \tilde{\Gamma}$ and
energies $\sigma \omega, \sigma\tilde{\omega}$ of these mesons.  (In
the $\Upsilon(4S)$ rest frame $B_{H}$ and $B_{L}$, which are produced
back-to-back, have (nearly) identical velocities and in this frame we
have $\Gamma_{H} = \tilde{\Gamma}_{H}, \Gamma_{L} =
\tilde{\Gamma}_{L},\sigma\omega = \sigma\tilde{\omega}$.)  Thus
\eqref{eq:20} becomes
\begin{equation}
  \label{eq:22}
  \vert S_{\mathcal{O}\tilde{\mathcal{O}}}\vert^{2} =
  \frac{1}{2}e^{-\Gamma t -\tilde{\Gamma}\tilde{t}}
  \Bigl\{
    \vert A_{L} \vert^{2}\vert \tilde{A}_{H} \vert^{2}  + 
    \vert A_{H} \vert^{2}\vert \tilde{A}_{L} \vert^{2}  
    - A_{L} A_{H}^{*}\tilde{A}_{H}\tilde{A}_{L}^{*} e^{i\xi}
    - A^{*}_{L} A_{H}\tilde{A}^{*}_{H}\tilde{A}_{L}e^{-i\xi}
   \Bigr\}  
\end{equation}
where
\begin{equation}
  \label{eq:25}
  \xi \equiv \delta
  m^{2}(\tilde{t}/\sigma\tilde{\omega}-t/\sigma\omega)\ .   
\end{equation}
We can equally well express $\xi$ in terms of the laboratory frame
distances the $B$ mesons travel, $d, \tilde{d}$: $\xi = \delta
m^{2}(\tilde{d}/\sigma\tilde{p} - d/\sigma p)$.  Alternatively, we may
use the $B$ meson decay times $T, \tilde{T}$ evaluated in the
$\Upsilon(4S)$ rest frame, where $\sigma\omega = \sigma\tilde{\omega}
\simeq \sigma m$ and $\xi = \delta m (\tilde{T}-T)$.  To simplify our
results we have ignored the difference in widths between the heavy and
light $B$ mesons, taking $\Gamma_{L} = \Gamma_{H} \equiv \Gamma$.
Only slightly more effort is required to keep track of this effect.

As our first example we evaluate the mixing probability obtained from
measurements in which we observe one $B$ meson decaying into a final
state $\ket{\mathcal{O}^{\pm}}$ and the other into a final state
$\ket{\tilde{\mathcal{O}}^{\pm}}$, each containing a charged lepton.
Because there is negligible direct $CP$ violation in these $B$ decays
the various amplitudes are related.  The $B_L$ meson in \eqref{eq:19}
may be created by a local operator of the form $p(\bar{d}b) +
q(\bar{b}d)$ where $p$ and $q$ are constants determined by the
requirement that this operator does not also create the $B_H$
meson. This leads to the usual expressions for $p$ and $q$ (with the
usual phase freedom). A similar argument shows that the operator $ p
(\bar{d}b) - q (\bar{b}d)$ creates only the $B_H$ meson. Imposing
$CP$ invariance in the time-development of the operators ${\cal
  O}^\pm$, $(CP) {\cal O}^\pm (CP) = {\cal O}^\mp$, yields the
relations:
\begin{equation}
  \label{eq:23}
  \begin{split}
  A_{L}^{+} = A_{H}^{+} = p A  &\qquad  A_{L}^{-}
  = -A_{H}^{-} = q A  \\ 
   \tilde{A}_{L}^{+} = \tilde{A}_{H}^{+}
  = p \tilde{A}
  &\qquad \tilde{A}_{L}^{-} = -\tilde{A}_{H}^{-}
  = q \tilde{A} \ .
  \end{split}
\end{equation}
Using these relations in \eqref{eq:22} we obtain
\begin{equation}
  \label{eq:24}  
\begin{split}
  \vert S_{++}\vert^{2} &= \vert A\vert^{2}
  \vert \tilde{A} \vert^{2}e^{-\Gamma t - \tilde{\Gamma}\tilde{t}}
  \vert p \vert^{4} 
  \sin^{2}\frac{\xi}{2} \\ 
  \vert S_{--}\vert^{2} &= \vert A\vert^{2}
  \vert \tilde{A} \vert^{2}e^{-\Gamma t - \tilde{\Gamma}\tilde{t}}
  \vert q \vert^{4} 
  \sin^{2}\frac{\xi}{2} \\
  \vert S_{-+}\vert^{2} &= \vert A\vert^{2}
  \vert \tilde{A}\vert^{2}e^{-\Gamma t - \tilde{\Gamma}\tilde{t}} \vert p \vert
  ^{2}\vert q \vert ^{2} \cos^{2} \frac{\xi}{2} \ .
\end{split}
\end{equation}

In the absence of $CP$ violation $\vert q/p\vert=1$. 
The mixing probability $\chi$ is then
\begin{equation}\label{mixrate}
\chi \equiv \frac{\iint_0^\infty \!dt d\tilde{t} (|S_{++}|^2+|S_{--}|^2)/2 }
  {\iint_0^\infty \! dt d\tilde{t}(|S_{++}|^2/2+|S_{--}|^2/2+|S_{-+}|^2)}
= \frac{x^2}{2(1+x^2)} \,,
\end{equation}
where $x \equiv \delta m^{2}/(\sigma\omega\,\Gamma)$ and we have
divided by $2$ when integrating to avoid double counting identical
final states. In evaluating this integral we have used the fact that
$\Gamma\,\sigma\omega$ is Lorentz invariant so that
$\Gamma\,\sigma\omega = \tilde{\Gamma}\,\sigma\tilde{\omega}$.
Further, this Lorentz invariance allows the evaluation of $x$ in the
rest frame where $\delta m^{2}/\sigma\omega =
(M^{2}_{H}-M^{2}_{L})/(M_{H}+M_{L}) = \delta m$ and $\Gamma =
\Gamma_{0}$. Therefore $x = \delta m/\Gamma_{0}$, and $\chi$ is seen
to be the usual expression.  If the difference in widths of the two
states is taken into account, we obtain $\chi =
(x^2+(\delta\Gamma/2\Gamma)^{2})/(2(1+x^2))$.

We turn to the time-dependent $CP$ asymmetries in $B$ meson decay. In
this case we tag one of the $B$ mesons via a decay to a charged lepton
as before, but then observe the decay of the other $B$ into a $CP$
eigenstate, $f$. Our previous analysis continues to apply: the $A$
amplitudes continue to refer to measurements involving a charged
lepton and are still given by \eqref{eq:23}.  The other amplitudes,
now denoted as $A^{f}_{L,H}$, refer to the detection of a $CP$
eigenstate $f$.

The two amplitudes $A^{f}_{L},A^{f}_{H}$ are in general independent. It is
conventional to define
\begin{equation}
A^f_{L,H} \equiv p A_f \pm  q \bar{A} _f = p A_f
(1 \pm \lambda_f)\,, 
\end{equation}
where $\lambda_f = (q/p)(\bar{A}_f / A_f)$.  The tagged rates are 
\begin{equation}
\label{timedep}
\begin{split}
  |S_{+f}|^2   & \propto e^{-\Gamma t -\tilde{\Gamma}\tilde{t}}\,
     |p^2 A A_f|^2 \bigg[ |\lambda_f|^2
  \cos^2\frac{\xi}{2} + \sin^2\frac{\xi}{2}
  + \text{Im}\lambda_f \sin\xi\bigg] \\
  |S_{-f}|^2  & \propto e^{-\Gamma t - \tilde{\Gamma}\tilde{t}}\,
   |p q A  A_f|^2 \bigg[ \cos^2\frac{\xi}{2} + |\lambda_f|^2
  \sin^2\frac{\xi}{2} - \text{Im}\lambda_f \sin\xi\bigg]\ ,
\end{split}
\end{equation}
where the tilded quantites refer to the observation of the $B$ meson
decaying into the $CP$ eigenstate $f$ and $\xi$ is  given by
\eqref{eq:25}. 
For the $B$ meson system $|p|\simeq
|q|$ and the time-depenent asymmetry is
\begin{equation}\label{acp}
a_f   \equiv \frac{|S_{+f}|^2 - |S_{-f}|^2}
  {|S_{+f}|^2 + |S_{-f}|^2} 
= \frac{(|\lambda_f|^2-1) \cos\xi
  + 2\,{\rm Im}\,\lambda_f \sin\xi}{1+|\lambda_f|^2} \,.
\end{equation}
Using the value of $\xi$ in the $\Upsilon(4S)$ rest frame this is seen
to be the standard expression~\cite{Harrison:1998yr}.

Other examples of neutral meson mixing may be treated similarly. The
universal formula \eqref{eq:11} allows a ready treatment of all
pertinent cases.

\section{Conclusions}
\label{sec:conclusions}

Oscillation phenomena, whether involving neutral mesons or neutrinos,
have been widely studied experimentally. 
Although attempts to describe the underlying theoretical formalism are
rife in the literature, the arguments used are often obscure,
confusing, or simply wrong. The starting point of many such analyses
is a ``flavor eigenstate'' which is neither an energy eigenstate nor
takes into account the entanglement of the neutrino with other final
state particles. This leads to equal-momentum versus equal-energy
controversies, to inappropriate appeals to ``energy--time
uncertainty,'' and to alleged ambiguities related to the oscillation
phase that are somehow to be resolved by future experiments.

In this paper, we present a theoretical analysis of oscillation
phenomena in a fashion that is simple, entirely general, and free of
ambiguities.  The oscillation phase is unambiguously given by
\eqref{eq:11}, an expression equally applicable for neutrinos and
mesons of any energy, relativistic or not.  The occurrence of
oscillations requires simply that 1.) the oscillating particles be
observed and 2.) condition \eqref{eq:9} be satisfied thus ensuring the
overlap of these particles at the time of their detection.

Our approach to oscillations shows that the variations in decay times
observed in the GSI experiment (where neutrinos from electron capture
are not observed) cannot be attributed to neutrino
mass mixing. Furthermore, we point out that our universal criterion is
satisfied by the proposed Raghavan experiment which, if it proves
feasible, should enable the observation of neutrino oscillations.

\begin{acknowledgments}
  We thank Stuart Freedman, Yuval Grossman, Witek Skiba and Jesse
  Thaler for helpful discussions, and Howard Haber and Bob Cahn for
  bringing the work of Nauenberg to our attention. We thank the Aspen
  Center for Physics where this work was begun. This work was
  supported in part by the U.S.\ Department of Energy under grants
  DE-FG02-01ER-40676 and DE-AC02-05CH11231.
\end{acknowledgments}

\bibliography{neutrinos}
\bibliographystyle{utphys}

\end{document}